# Current-voltage characteristics of NdFeAsO$_{0.85}$F$_{0.15}$ and NdFeAsO$_{0.85}$ superconductors

Yong Liu, Yi Sheng Chai, Hyeong-Jin Kim, G. R. Stewart[†], and Kee Hoon Kim[*]

*Department of Physics and Astronomy, Seoul National University, Seoul 151-747, Republic of Korea*

Zhi-An Ren and Zhong-Xian Zhao

*National Laboratory for Superconductivity, Institute of Physics and Beijing National Laboratory for Condensed Matter Physics, Chinese Academy of Sciences - P. O. Box 603, Beijing 100190, People's Republic of China*

**Abstract**

The vortex phase diagrams of NdFeAsO$_{0.85}$F$_{0.15}$ and NdFeAsO$_{0.85}$ superconductors are determined from the analysis of resistivity and current-voltage (*I-V*) measurements in magnetic fields up to 9 T. A clear vortex glass to liquid transition is identified only in the oxygen deficient NdFeAsO$_{0.85}$, in which *I-V* curves can be well scaled onto liquid and glass branches consistent with the vortex glass theory. With increasing magnetic field, the activation energy $U_0$, deduced from the Arrhenius plots of resistivity based on the thermally activated flux-flow model (TAFF), decays more quickly for NdFeAsO$_{0.85}$F$_{0.15}$ than for NdFeAsO$_{0.85}$. Moreover, the irreversibility field $H_{irr}$ of NdFeAsO$_{0.85}$ increases more rapidly than that of NdFeAsO$_{0.85}$F$_{0.15}$ with decreasing temperature. These observations evidence the strong vortex pinning effects, presumably caused by the enhanced defects and disorders in the oxygen deficient NdFeAsO$_{0.85}$. It is inferred that the enhanced defects and disorder can be also responsible for the vortex glass to liquid

---

[*] Corresponding author; khkim@phya.snu.ac.kr (K. H. Kim)

[†] On sabbatical from Department of Physics, University of Florida

transition in the NdFeAsO$_{0.85}$.

**I. INTRODUCTION**

The recently discovered iron-arsenide superconductors have attracted much interest due to their possible unconventional superconducting pairing mechanism and application aspects [1,2]. Representative systems so far found can occur either in the 1111 phase with the general formula ReFeAsO (Re: rare-earth elements) or in the 122 phase with the formula of $M$Fe$_2$As$_2$ ($M$=Ba, Sr, and Eu) [1-4]. In the former, the O$^{2-}$ ions can be substituted by F$^-$ ions to introduce hole carriers while in the latter, the substitution of K$^+$ for $M^{2+}$ can introduce hole carriers [3,4]. Both representative systems have characteristic layered structures that have stacking of ReO (or $M$) and FeAs layers along the $c$-axis, in which the ReO (or $M$) layer behaves like a charge reservoir block and the FeAs layer acts as the conducting layer.

Although lots of research efforts have been already made to identify the superconducting gap symmetry or superconducting pairing mechanism, there exist only a limited number of studies on the vortex behavior of these compounds. In a NdFeAsO$_{0.82}$F$_{0.18}$ single crystal, it was found the upper critical fields determined based on the Werthamer–Helfand–Hohenberg formula are $H_{c2}^{ab}(0) \approx 304$ T and $H_{c2}^{c}(0) \approx 62 \sim 70$ T [5]. These high upper critical fields promise potential application of the new superconductors in carrying strong current and producing high magnetic field.

Based on heat capacity measurements of single crystal of NdFeAsO$_{1-x}$F$_x$, the zero-temperature coherence lengths of $\xi_{ab}(0)$ and $\xi_c(0)$ were estimated to be 3.7 nm and 0.9 nm, respectively, resulting in an anisotropy of $\xi_{ab}(0)/\xi_c(0) \cong 4$ [6]. We note that these properties of the layered structure (small coherence lengths and moderate anisotropy) in the iron-arsenide superconductors are reminiscent of high-$T_c$ cuprate superconductors (HTS) [7], where the coherence lengths are rather small so that the point defects and dislocations can serve as pinning centers. Therefore, it might be expected that a rich vortex phase diagram can also exist in the iron-arsenide superconductors just as in the HTS.

In this report, the vortex phase transition boundaries of $NdFeAsO_{0.85}F_{0.15}$ and $NdFeAsO_{0.85}$ superconductors are investigated by analyzing the irreversibility line (IL) and current - voltage (*I-V*) curves. In the vortex liquid phase of both compounds, it is found that the resistivity tail can be described by the thermally activated flux-flow (TAFF) model [8,9]. *I-V* measurement across the IL reveals that a vortex liquid to glass transition only exists in the oxygen-deficient $NdFeAsO_{0.85}$ sample.

## II. EXPERIMENTAL

Polycrystalline samples of $NdFeAsO_{0.85}F_{0.15}$ and $NdFeAsO_{0.85}$ were prepared by the quartz-tube-sealing and high pressure synthesis methods, respectively. The detailed procedures for sample preparation have been reported elsewhere [10-12]. Four-probe resistivity measurements under magnetic fields up to 9 T were performed using an AC resistance bridge in a Physical Property Measurement System (PPMS[TM], Quantum Design). *I-V* curves were measured with a DC current source (Keithley 236) and a nanovoltmeter (Keithley 182).

## III. RESULTS AND DISCUSSION

Figures 1 (a) and (c) show the resistivity broadening of $NdFeAsO_{0.85}F_{0.15}$ and $NdFeAsO_{0.85}$ with increasing magnetic field. Both samples show a superconducting transition around 45 K. With the application of magnetic field the resistivity drop significantly broadens for both compounds, and displays a long tail that merges into a flat floor dominated by noise. The TAFF model in the vortex liquid state can explain the resistivity tail in applied magnetic field [8,9]:

$$\rho(H,T) = \rho_0 \exp\{-U_0 / k_B T\},  \tag{1}$$

where $U_0$ corresponds to the activation energy for the vortex motion.

Figures 1 (b) and (d) present the Arrhenius plots of the experimental data shown in Figs. 1 (a) and (c). The field dependence of $U_0$ can be deduced from the linear fit to each curve, as indicated by the dashed lines in the semilog plots of Figs. 1 (b) and (d). The thus obtained field dependence of the activation energy $U_0(H)$ is summarized in Fig. 2. Following the HTS case, we use a power law [13,14]

$$U_0(H) \propto H^{-\alpha} \qquad (2)$$

to describe $U_0(H)$ for NdFeAsO$_{0.85}$F$_{0.15}$ and NdFeAsO$_{0.85}$. The double-logarithmic plot of $U_0(H)$ in Fig. 2 shows two linear regimes for both samples with a kink around 1 T. From the linear fit to the curve, we find that for NdFeAsO$_{0.85}$F$_{0.15}$, $\alpha \approx 0.22$ in the low field region below $H=1$ T and $\alpha \approx 0.68$ in the high field region $H > 1$ T. A similar behavior is also observed in NdFeAsO$_{0.85}$; the linear fits produce $\alpha \approx 0.20$ for $H < 1$ T and $\alpha \approx 0.65$ for $H > 1$ T. In a similar sample NdFeAsO$_{0.82}$F$_{0.18}$, a roughly linear behavior with $\alpha \approx 1/3$ is observed within the field range $0 < H < 13$ T [15].

We note that the $U_0$ of NdFeAsO$_{0.85}$F$_{0.15}$ decreases more quickly than that of NdFeAsO$_{0.85}$ with increasing magnetic fields. As shown in Fig. 2, the field dependences of $U_0$ for the two samples almost cross in the high field regime. For the oxygen deficient NdFeAsO$_{0.85}$, the vortex pinning can be enhanced because the oxygen deficiency in NdFeAsO$_{0.85}$ can cause disorder and introduce point-like defects. The effect of randomly distributed oxygen is directly reflected in the superconducting transition width $\Delta T_c$. As shown in Figs 1 (a) and (c), in zero field, a larger $\Delta T_c$ (~ 7 K) is observed in NdFeAsO$_{0.85}$ than in NdFeAsO$_{0.85}$F$_{0.15}$ ($\Delta T_c \sim 3$ K), indicating a broader $T_c$ distribution. With increasing field, weak superconducting regions due to the fluctuated oxygen content will be driven to the normal state to become effective pinning centers. According to the theory of collective pinning [16], the point defects have a significant influence on the characteristic length $L_c$ of vortex loops.

We define the irreversibility temperature $T_{irr}(H)$ with the zero resistance criterion of 0.01 mΩ cm, which is close to the noise floor of the resistance measurements. Plotting the irreversibility limits $T_{irr}(H)$ of the both samples for various applied fields in the $H$-$T$ diagram generates an IL that separates the $H$-$T$ plane into two regions, i.e., vortex liquid and solid regions, as shown in Fig. 3. The upper critical fields are determined from the mean-field critical temperature $T_c(H)$ of the samples, which corresponds to the peak value in the plots of $d\rho(H)/dT$ vs $T$.

In general, the IL is described by a power law [13,17]

$$H_{irr}(T) = H_0[1 - T_{irr}(H)/T_{irr}(0)]^n, \qquad (3)$$

where $H_{irr}(T)$ is the irreversibility field as a function of the temperature, and $n$ and $H_0$ are the fitting parameters. The fits to the IL yield $n \sim 1.4$ and 1.9 for $NdFeAsO_{0.85}F_{0.15}$ and $NdFeAsO_{0.85}$, respectively. The value of exponent $n$ is governed by the physics of pinning potential and dimensionality in superconductors [18]. It is clearly seen in Fig. 3 that $H_{irr}(T)$ of $NdFeAsO_{0.85}$ increases more quickly with decreasing temperature compared to that of $NdFeAsO_{0.85}F_{0.15}$, as is also clear from the larger exponent of 1.9. This feature is consistent with the field-dependence of activation energy $U_0$. The strength of vortex pinning is enhanced for $NdFeAsO_{0.85}$ in the low temperature and high field regime.

In order to study the nature of the vortex phase transition in superconducting $NdFeAsO_{0.85}F_{0.15}$ and $NdFeAsO_{0.85}$, $I$-$V$ measurements were performed across the IL determined in Fig. 3. Figures 4 (a)-(d) summarize the thus calculated $E$–$J$ curves measured at 35 K and 37 K under various magnetic fields. Here, the electric field $E$ is calculated with the relation $E = V/L$, where $V$ is the measured voltage and $L$ is the distance between the two voltage contacts. The current density $J$ is determined using the relation $J = I/A$, where $A$ is the cross-section area of the sample. For $NdFeAsO_{0.85}F_{0.15}$, a linear behavior with slope ~ 1 is observed in the small current regime (Figs. 4(a) and (c)). With decreasing magnetic field, the *E-J* curves shift toward the large current regime almost in parallel. In sharp contrast to this behavior in NdFeAsO$_{0.85}$F$_{0.15}$, *E-J* curves in NdFeAsO$_{0.85}$ begin to deviate from the linear behavior below a characteristic field $H_g$, displaying a negative curvature (Figs. 4(b) and (d)). At the fixed *H*= 9 T, *E-J* curves in NdFeAsO$_{0.85}$F$_{0.15}$ (Fig. 4(e)) show a strong upward behavior at high temperatures and transform to a linear behavior at low temperature. For NdFeAsO$_{0.85}$, however, a negative curvature is clearly observed below a characteristic temperature $T_g$.

The linear behavior with slope of ~1 implies an ohmic dissipation in the vortex liquid regime. Within the TAFF model [8,9], a general expression for the *E-J* relation is

$$E = 2\nu_0 BL \exp(-U_0/k_B T) \sinh(JBV_C L_p/k_B T), \qquad (4)$$

where $U_0$ is the activation energy, $L_p$ is the pinning potential range, and $\nu_0$ is the attempt frequency of a flux bundle of volume $V_C$ to hop a distance *L*. For small driving current *J* in which $JBV_C L_p \leq k_B T$, $\sinh(x) \simeq x$. Thus we have

$$E = 2\nu_0 L \frac{JB^2 V_C L_p}{k_B T} \exp(-U_0/k_B T). \qquad (5)$$

Therefore, electric field *E* is independent of current density *J*, which predicts an ohmic dissipation at the low *J* limit. This is consistent with the slope of ~ 1 observed in the ohmic dissipation regime in Figs. 4 (a) and (b). With increasing current density *J* when $JBV_C L_p \gg k_B T$, the sinh term of Eq. (5) will prevail, and lead to exponential behavior at the large current regime, as observed in Fig 4(e).

On the other hand, NdFeAsO$_{0.85}$ is characterized by negative curvature below $H_g$ or $T_g$. According to the vortex glass theory [19,20], the *J* dependence of *E* at temperatures $T < T_g$ is described as

$$E \propto \exp[-(J_c/J)^\mu], \qquad (6)$$

where $\mu$ is a universal exponent. This equation shows negative curvature in a double-logarithmic plot. In a 3D case, one can scale E–J curves in the form [21,22]

$$\tilde{E} \sim E/(J \times |T-T_g|^{\nu(z-1)}) \text{ and } \tilde{J} \sim J/|T-T_g|^{2\nu},\qquad(7)$$

where $\nu$ and $z$ correspond to the correlation length and dynamical critical exponents, respectively. According to the theory, the E-J curves above and below $T_g$ will collapse onto two branches, corresponding to the vortex liquid and glass states, respectively. Figure 5(a) shows that indeed the E-J data scale onto two curves (T above and below $T_g$) with the scaling exponents $\nu = 1.8$ and $z = 1.4$.

As a further check of the validity of this theory in describing our data, in addition to the scaling behavior of E–J curves, the vortex glass theory also expects that, approaching the critical regime, resistivity should vanish as [19-21]

$$\rho \propto (T-T_g)^{\nu(z-1)}.\qquad(8)$$

Therefore, a plot by transforming Eq. (8) as

$$(d\ln\rho/dT)^{-1} \propto \frac{1}{\nu(z-1)}(T-T_g)\qquad(9)$$

should result in a straight line with intercept $T_g$ and slope $1/\nu(z-1)$ extrapolating $(d\ln\rho/dT)^{-1}$ to 0 [23]. A linear behavior of $(d\ln\rho/dT)^{-1}$ vs $(T-T_g)$ with slope of $1.26\pm0.30$ is indeed observed in resistivity data obtained at 9 T just above $T_g$, as shown in Fig. 5(b), which is consistent with the scaling exponents ($1/\nu(z-1) = 1.39$) determined from the E-J curves in Fig. 5a and discussed above.

How do the values determined for $\nu$ and $z$ compare with those found in other superconductors? It should be stressed that in the Ising spin glass one expects $z > 4$ and $\nu \sim 1 - 2$ in the 3D vortex glass transition [20,21], consistent with the most cases in HTS. In the polycrystalline specimens of $YBa_2Cu_3O_{7-\delta}$, however, the values of $z$ were found to vary within the range $2.3 < z < 4.3$, being field- and morphology-dependent [24-26].

Therefore, the small dynamical critical exponent *z* observed here might be also caused by the granular nature of the samples. It was suggested that superconducting properties of $YBa_2Cu_3O_{7-\delta}$ polycrystalline sample can be explained within the framework of the gauge glass model [24], where the small dynamical critical exponent *z* was predicted [27,28].

A question then arises why we cannot observe the vortex glass phase transition in $NdFeAsO_{0.85}F_{0.15}$. A simple explanation is that the collective-creep theory considers the case where the randomness is weak enough ($L_c \leq L_{LO}$) to enable a description of the local vortex lattice in terms of elastic-media theory, whereas the vortex glass theory considers long length scales ($L_c \gg L_{LO}$) in which $L_c$ and $L_{LO}$ reflect the characteristic length of vortex loops and the size of the "Abrikosov lattice domains", respectively. Thus, the vortex glass behavior will be observed more easily in more strongly disordered materials with small $L_{LO}$ [28]. The obvious difference between $NdFeAsO_{0.85}F_{0.15}$ and $NdFeAsO_{0.85}$ is that there exist large amounts of oxygen deficient defects and disorder in $NdFeAsO_{0.85}$, which is a likely explanation for the occurrence of vortex glass transition.

Finally, we would like to point out that the IL of $NdFeAsO_{0.85}F_{0.15}$ corresponds to a crossover line. Above the IL, the plastic creep of flux lines the can be greatly enhanced due to the thermal activation and below the IL, a few flux lines can still move without achieving a real zero resistance. In Fig. 3, we find that the IL line overlaps with the characteristic $H_g$ and $T_g$ (denoted by stars in Fig. 3), constituting strong evidence that the IL is the vortex glass to liquid melting line in $NdFeAsO_{0.85}$.

## IV. CONCLUSION

In summary, we have studied the resistive broadening behavior and *I-V* characteristics in $NdFeAsO_{0.85}F_{0.15}$ and $NdFeAsO_{0.85}$ superconductors. The field-dependence of activation energy $U_0(H)$ and temperature-dependence of irreversibility field $H_{irr}(T)$ suggest that the randomly distributed weak links due to oxygen deficiency in NdFeAsO$_{0.85}$ become pinning centers in the low temperature and high field regime. *I-V* measurements suggest that a vortex liquid to glass transition occurs in the sample NdFeAsO$_{0.85}$. The irreversibility line of NdFeAsO$_{0.85}$ determined from the resistivity measurements coincides with the vortex liquid to glass transition.

## ACKNOWLEDGMENTS

We appreciate helpful discussions with Prof. S. I. Lee. This work was supported by the National Research Lab program (M10600000238) and KRF Grant (KRF-2008-314-C00101). Work at Florida supported by the US Department of Energy, contract no. DE-FG02-86ER45268.

**Figure captions**

Fig. 1. (Color online) Left panel shows the resistivity curves for (a) NdFeAsO$_{0.85}$F$_{0.15}$ and (c) NdFeAsO$_{0.85}$ under various applied magnetic fields. Corresponding Arrhenius plots (log $\rho$ vs 1/$T$) were plotted in the right panel for (b) NdFeAsO$_{0.85}$F$_{0.15}$ and (d) NdFeAsO$_{0.85}$. Dashed lines are the linear fitting results.

Fig. 2. (Color online) The field dependence of the activation energy $U_0$ for NdFeAsO$_{0.85}$F$_{0.15}$ and NdFeAsO$_{0.85}$ in double-logarithmic plot (Eq. (2)). Solid lines are linear fits.

Fig. 3. (Color online) The vortex phase diagrams of NdFeAsO$_{0.85}$F$_{0.15}$ and NdFeAsO$_{0.85}$ superconductors with the irreversibility lines and upper critical field included. The solid lines are fittings to the power-law (1-$T_{irr}(H)/T_{irr}(0)$)$^n$, which yields $n \sim$ 1.4 and 1.9 for NdFeAsO$_{0.85}$F$_{0.15}$ and NdFeAsO$_{0.85}$ superconductors, respectively. Green stars correspond to the glass transition temperature (or field) $T_g$ ($H_g$).

Fig. 4. (Color online) Double-logarithmic plots of electric field $E$ vs current density $J$ curves at 35 K and 37 K for NdFeAsO$_{0.85}$F$_{0.15}$ ((a) and (c)) and NdFeAsO$_{0.85}$ ((b) and (d)) under different magnetic fields. The blue lines in (a) and (b) correspond to a slope of 1. (e) and (f) $E$-$J$ curves at 9 T for temperatures from 25 K to 33 K with an interval 0.5 K. Dashed lines in (b), (d) and (f) indicate the vortex glass transition field $H_g$ and temperature $T_g$, where a negative curvature is observed below $H_g$ or $T_g$.

Fig. 5. (Color online) (a) *E-J* curves shown in Fig. 4(f) collapse into two branches, above and below $T_g$, by scaling as $\tilde{E} \sim E/(J \times |T-T_g|^{\nu(z-1)})$ and $\tilde{J} \sim J/|T-T_g|^{2\nu}$. (b) Plot of the inverse logarithmic derivative of resistivity at 9 T. Red line is a linear fit with slope of $1.26 \pm 0.30$. The arrow indicates the vortex glass transition temperature $T_g = 32$ K.

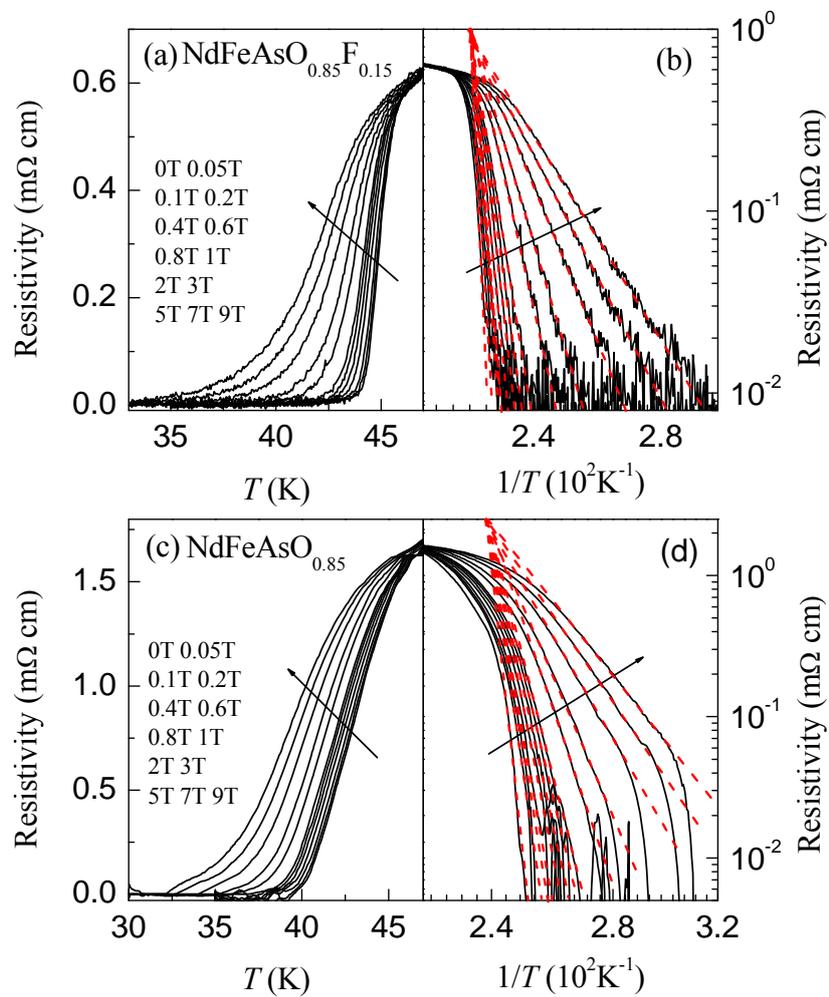

Fig. 1 Yong Liu *et al*.

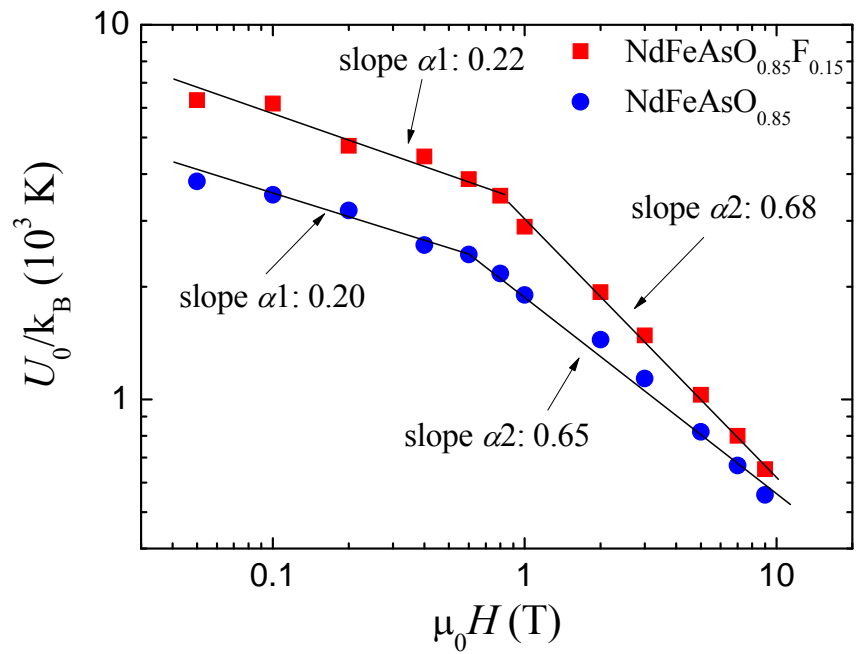

Fig. 2 Yong Liu *et al*.

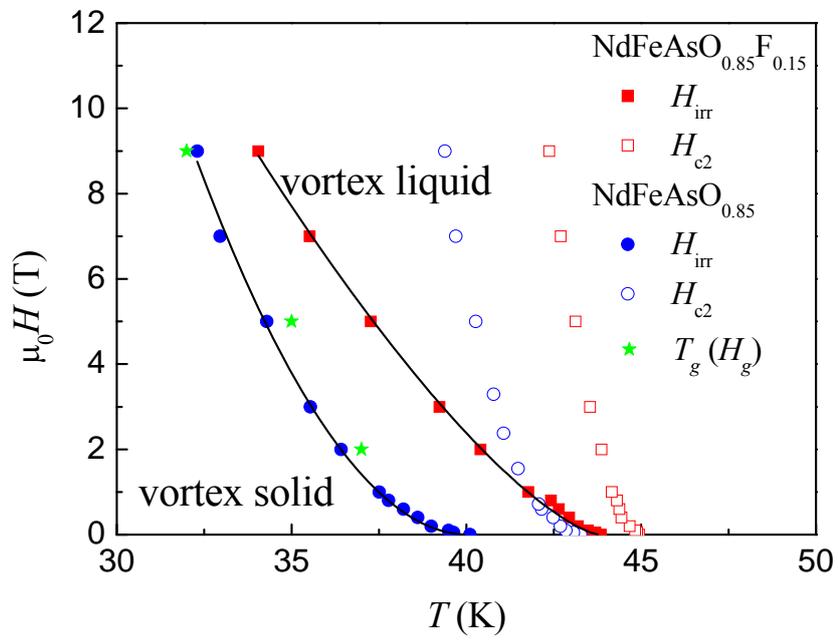

Fig. 3 Yong Liu *et al*.

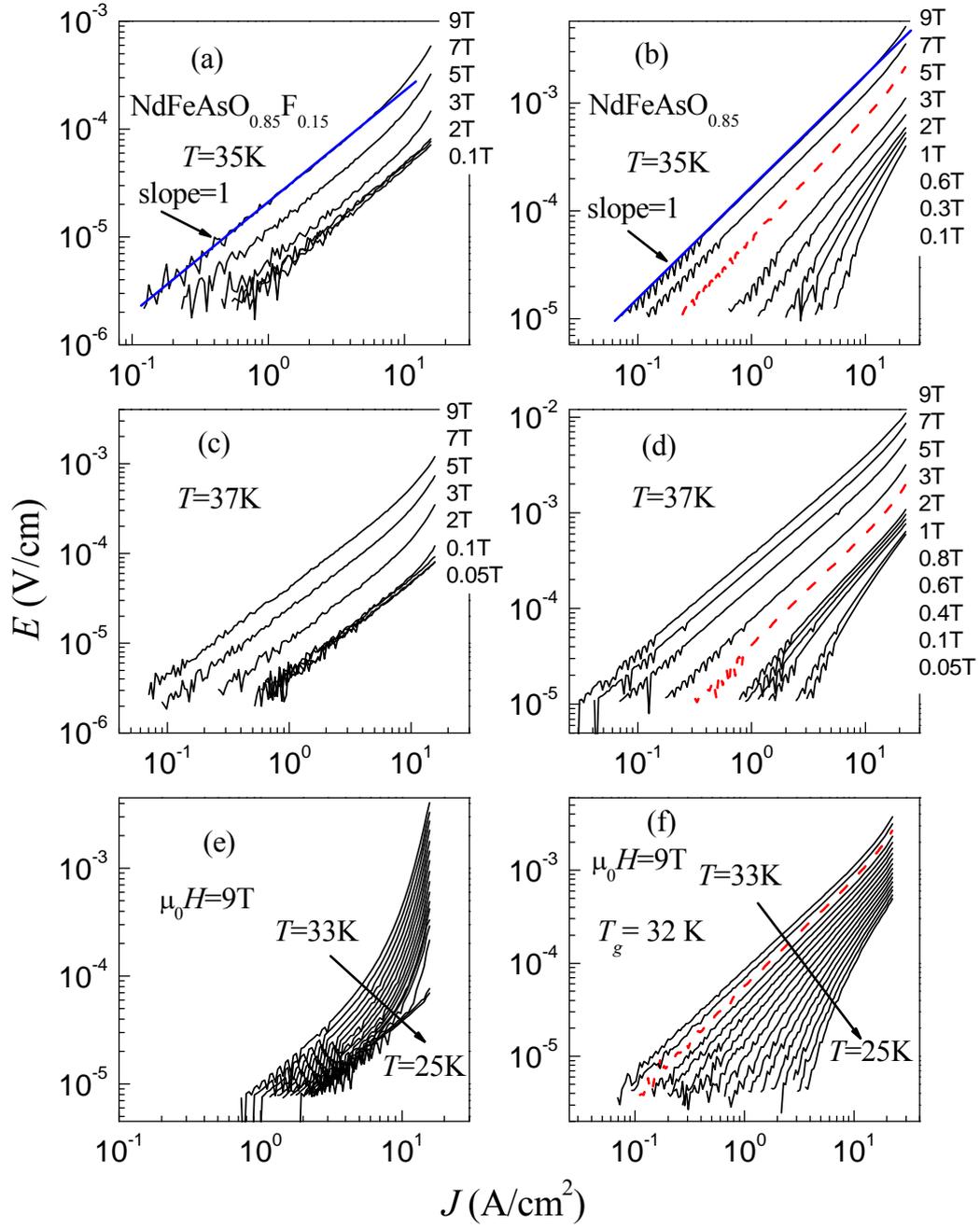

Fig. 4 Yong Liu *et al*.

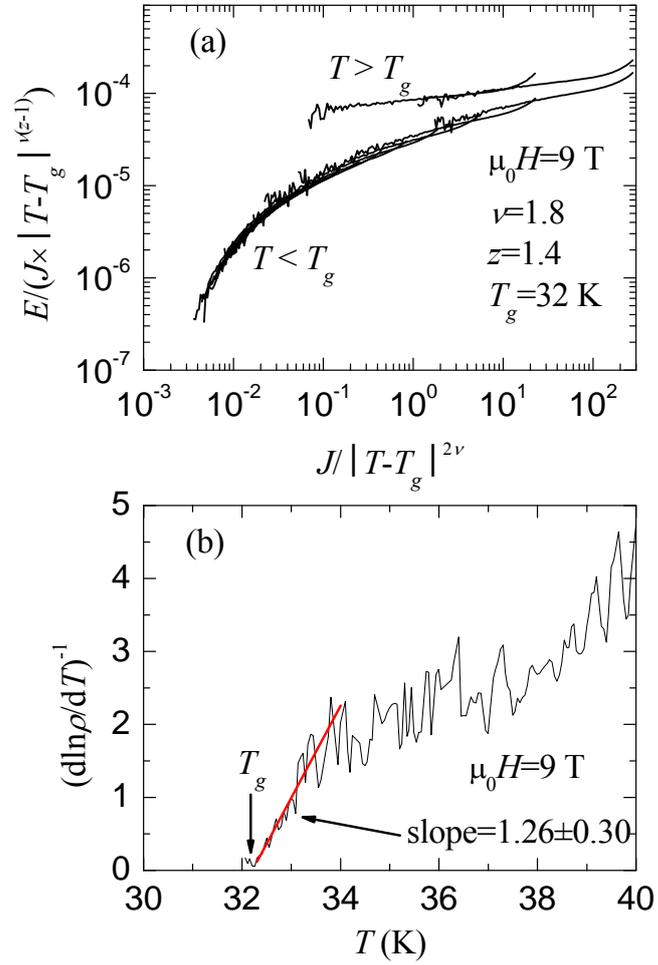

Fig. 5 Yong Liu *et al*.